\documentclass[a4paper,12pt]{article}
\usepackage[utf8]{inputenc}

\usepackage{amsmath,amsfonts,amsthm,amssymb,dsfont}
\usepackage{graphicx,wrapfig,lipsum}
\usepackage[section]{placeins}
\usepackage{braket}

\usepackage[numbers,sort&compress]{natbib}
\usepackage{mathrsfs}
\usepackage{tikz}
\usetikzlibrary{shapes.geometric,decorations.markings}
\usepackage{subfigure}

\global\long\def\SU{\mathrm{SU}}%
\theoremstyle{plain}
    \ifx\thechapter\undefined
      \newtheorem{prop}{\protect\propositionname}
    \else
      
    \fi
\providecommand{\propositionname}{Proposition}

\usepackage[hidelinks]{hyperref}

\hypersetup{
    colorlinks=true,
    linktocpage=true,
    linkcolor=blue,
    urlcolor=blue,
    citecolor=blue,
}

\numberwithin{equation}{section}
\newcommand{\Vol}{\text{Vol}}
\newcommand{\vol}{\text{vol}}

\newcommand{\dd}{\mathrm{d}}
\newcommand{\AdS}[1]{\text{AdS}_{#1}}
\newcommand{\Sp}[1]{\text{S}^{#1}}
\renewcommand{\AA}{\mathrm{A}}
\newcommand{\ellAdS}{\ell_{\text{AdS}_{3}}}
\newcommand{\ellS}{\ell_{\text{S}^{2}}}

\renewcommand{\SU}[1]{\text{SU}(#1)}
\newcommand{\ri}{\mathrm{i}}

\newcommand{\ds}{\dd s^{2}}
\newcommand{\hypergeomF}[4]{{}_{2}F_{1}\left(#1,#2;#3;#4\right)}

\begin{document}

\begin{titlepage}

\begin{center}

$\phantom{.}$\\ \vspace{2cm}
\noindent{\Large{\textbf{(Iso)spin from Isospin in Top-Down Holography}}}

\vspace{1cm}

Marcelo Oyarzo$^{\text{a,b}}$ \footnote{moyarzoca1@gmail.com}, 
Ricardo Stuardo$^{\text{c,d}}$ \footnote{ricardostuardotroncoso@gmail.com}

\vspace{0.5cm}

$^\text{a}${Departamento de F\'isica de Part\'iculas \& Instituto Galego de Física de Altas Enerx\'ias (IGFAE), Universidade de Santiago de Compostela, E-15782 Santiago de Compostela, Spain}
		\\
\vspace{0.5cm}
$^\text{b}$Dipartimento di Fisica, Politecnico di Torino,\\
Corso Duca degli Abruzzi 24, I-10129 Turin, Italy and INFN, Sezione di Torino, Italy\\
\vspace{0.5cm}
$^{\text{c}}$Departamento de Física, Universidad de Oviedo,\\
Avda. Federico García Lorca 18, 33007 Oviedo, Spain\\
\vspace{0.5cm}
$^{\text{d}}$Instituto Universitario de Ciencias y Tecnologías Espaciales de Asturias (ICTEA),\\ Calle de la Independencia 13, 33004 Oviedo, Spain

\end{center}

\vspace{0.5cm}

\centerline{\textbf{Abstract}} 

\vspace{0.5cm}

\noindent{Motivated by the spin from isospin mechanism of Jackiw-Rebbi-Hasenfratz-’t Hooft, we study two SU(2) gauged supergravity solutions of the form $M_{d}\times \text{S}^{2}$ containing non-Abelian hedgehog  monopole on the 2-sphere. Due to the presence of the monopole, the SO(3) isometry group of the 2-sphere is not a symmetry of the configuration. Instead, a diagonal combination of the SU(2) gauge and the SO(3) isometry of the 2-sphere is the true symmetry of the configuration. Uplifting the solutions to Type II, the gauge-isometry diagonal symmetry becomes a diagonal combination between the SO(3) symmetry of the 2-sphere and a SU(2) symmetry of a 3-sphere used to uplift the configuration. One of the uplifts is supersymmetric and corresponds to the I-brane theory on a 2-sphere. The second background is a deformation of AdS$_{5}\times\Sp{5}$ and is not supersymmetric. We study dilaton fluctuations on the later geometry. Due to the diagonal symmetry, the fluctuations show angular momentum mixing between the SU(2) and SO(3) spins, mimicking the spin from isospin mechanism.
}

\vspace*{\fill}

\end{titlepage}

\newpage

\tableofcontents
\thispagestyle{empty}

\newpage
\setcounter{page}{1}
\setcounter{footnote}{0}

\section{Introduction}

String theory and the Gauge/Gravity duality \cite{Maldacena:1997re,Gubser:1998bc,Witten:1998qj,Itzhaki:1998dd,Boonstra:1998mp} have been fruitful for both sides of the duality. On one hand, the construction of Type II background have provided different insights on the field theory side. Some remarkable examples include the study of non-lagrangian field theories, such as the 6D $\mathcal{N}=(2,0)$ theory (see for example \cite{Henningson:1998gx,Beem:2014kka,Bullimore:2014upa}), and the study of confinement and phase transitions \cite{Witten:1998zw}. On the other hand, field theory ideas have also contributed to the development of gravitational solutions or methods to compute quantities holographically. For example, the Gaiotto-Maldacena backgrounds \cite{Gaiotto:2009gz} were constructed to holographically describe 4D $\mathcal{N}=2$ quiver field theories, while the Ryu–Takayanagi formula for entanglement entropy \cite{Ryu:2006bv} and holographic c-theorems \cite{Casini:2004bw,Myers:2010tj} are different tools whose construction was motivated by field theory. In this paper we follow the later approach,  focusing on a field theory result and use that as inspiration to look for supergravity solutions realising a similar phenomena.

Specifically, we focus on 4D SU(2) gauge theories in flat spacetime that allow for hedgehog magnetic monopoles. There is an interesting  phenomena mixing global and gauge symmetries \cite{Jackiw:1976xx,Hasenfratz:1976gr}: the non-Abelian field strength is not invariant under the SO(3) Lorentz symmetry, but this rather SO(3) invariant up to an SU(2) gauge transformation. This makes it so that the true symmetry of the configuration is a diagonal combination of the spacetime and gauge symmetry, and hence quantum fluctuations of fields around this configuration are to be classified in representations of this diagonal symmetry. 

Consider for example a scalar in a given representation of the gauge group. Fluctuations of this field appears as spin-0 with respect to the SO(3) symmetry, could actually have non-zero spin depending on its representation of the SU(2) gauge group, since the conserved charge that classifies the representation is the sum of the orbital SO(3) and internal SU(2) spin. Due to this feature, this mechanism is known as ``spin from isospin".  It was then suggested in \cite{Canfora:2012ap} that this mechanism could be used in the context of holography to obtain observables with fermionic quantum numbers, using only the naive bosonic fields. 

To this aim, we consider two gauged supergravities that contain at least an SU(2) factor as their gauge group. For the gauge field, we use the same configuration as in \cite{Canfora:2012ap}, which are called Merons: these are a simple non-Abelian configuration of the form $A = \lambda \hat{A}$, with $\hat{A}$ pure gauge and $\lambda \neq 1$. These configurations are then uplifted to 10-dimensions using internal manifolds than contain a 3-sphere. At this level, the gauge-isometry mixing becomes a diagonal combination of isometries of the external and internal manifolds, which is the closest one can get to the results of \cite{Jackiw:1976xx,Hasenfratz:1976gr} from holography\footnote{This is due to the fact that in a holographic set-up, one does not have access to the gauge symmetry of the dual theory, only to the global symmetries.}. A different approach to this problem can be found in \cite{David:2001nq}.

This paper is organized as follows: In Section \ref{sec:Section2} we review generic aspects of the geometries we consider and the Meron gauge field and its uplifts using 3-spheres. Section \ref{sec:Ibranes} studied the first of two configurations containing Meron gauge fields. In this case, the theory under consideration is the I-brane theory \cite{Itzhaki:2005tu} on $\Sp{2}$ with a SU(2) background Meron gauge field for the R-symmetry, which was found at the level of 4D gauged supergravity in \cite{Canfora:2021nca}. We show that the solutions preserves four supercharges for a choice of parameters, and that the R-symmetry of the configuration is in fact the diagonal symmetry. The second background configuration consists of a deformation of AdS$_{5}\times \Sp{5}$ and is studied in Section \ref{sec:Section3}. This solution is new, and is constructed at the level of the 5D SU(2)$\times$U(1) gauged supergravity. In Section \ref{sec:Section4} we study dilaton fluctations on the later background. We find a spectrum which shows angular momentum coupling between the symmetries of the configuration, mimicking the mechanism of \cite{Jackiw:1976xx,Hasenfratz:1976gr}. Section \ref{sec:Conclusions} contains a summary of our results, conclusions and ideas for future research.

\section{On the Meron Gauge Field and its Uplifts on \texorpdfstring{$\Sp{3}$}{S3}}\label{sec:Section2}

\subsection{SU(2) Meron Field}

In Yang-Mills theory, a Meron gauge field \cite{deAlfaro:1976qet, Eguchi:1980jx} is defined as 
    \begin{equation}
        A = \lambda U^{-1} \dd U.
    \end{equation}
Although this configuration is proportional to a pure gauge one, it is in fact the simplest intrinsically non-Abelian configuration. In fact, its field strength, $F = \dd A + g A\wedge A$ with $g$ the gauge coupling, is non-zero\footnote{We are in conventions where the coupling constant appears in the definition of the field strength instead of as a coefficient on the kinetic term. A similar, and more detailed, discussion in the other convention can be found in \cite{Canfora:2012ap}.}
    \begin{equation}
        F = \lambda\left( 1-g \lambda\,  \right) \dd U^{-1}\wedge \dd U,
    \end{equation}
so that this configuration is only pure gauge if $\lambda = g^{-1}$, as expected. In what follows we focus on SU(2) gauge theories, so that $U \in$ SU(2), and on spacetimes solutions of SU(2) gauged supergravities containing a 2-sphere, that is
    \begin{equation}\label{eq:MetricAnsatz}
        \dd s^{2} = g_{ab}\dd x^{a} \dd x^{b} + f(x) \dd s^{2}(\Sp{2})\,.
    \end{equation}
To parametrize the 2-sphere, it is convenient to use the embedding coordinates $\mu^{i}$, with $i=1,2,3$, satisfying $\mu^{i}\mu^{i}=1$, so that $\dd s^{2}(\Sp{2}) = \dd \mu^{i}\dd \mu^{i}$. This allows us to choose the following  hedgehog configuration for the SU(2) Meron\footnote{A more general ansatz can be chosen, but this is enough for our discussion.}
    \begin{equation}
        U = 2 \mu^{i} T^{i},
    \end{equation}
where the SU(2) generators are $T^{i} = -\frac{i}{2} \sigma^{i}$, with $\sigma^{i}$ the Pauli matrices, which sets the structure constants to $f_{ijk}=\epsilon_{ijk}$. With this ansatz, the gauge field and the field strength take the form
    \begin{equation}\label{eq:MeronFields}
        A^{i} = - 2\lambda \epsilon_{ijk}\mu^{j}\dd\mu^{k}, \quad
        F^{i} = -4\lambda (1-g\lambda) \mu^{i} \vol(\Sp{2}).
    \end{equation}

At this stage, it is already possible to see the mixing of the spacetime SO(3) with the gauge SU(2) symmetries. It is clear from \eqref{eq:MetricAnsatz} that the spacetime under consideration is invariant under the SO(3) isometry group of the 2-sphere. However, the field strength \eqref{eq:MeronFields} is not SO(3) invariant due to factor of $\mu^{i}$. Interestingly, under a simultaneous SO(3) and SU(2) action, the field strength is indeed invariant, revealing the true symmetry of the system. This SO(3) invariance up to a SU(2) transformation was discussed in \cite{Hasenfratz:1976gr,Jackiw:1976xx} for a different type of hedgehog configuration. There it was argued that since the conserved charge is the sum of spin $\vec{J}$ and isospin $\vec{T}$
    \begin{equation}\label{eq:MixingLorentzGauge}
        \vec{L}_{\text{total}} =\vec{J}  + \vec{T},
    \end{equation}
one should classify quantum fluctuations with respect to the quantum numbers of this symmetry. Hence, a naive SO(3) boson could actually have fermionic quantum numbers depending on its SU(2) representation. It is important to also remark that the components of the Meron gauge field \eqref{eq:MeronFields} are 1-forms dual to the $\Sp{2}$ Killing vectors.

In the following sections, we will work on gauged supergravities where the equation of motion of the non-Abelian gauge field reduces to
    \begin{equation}
        \dd \star F + g \left[A , \star F \right] = 0
        \quad  \Rightarrow \quad
        \lambda = \frac{1}{2g} 
    \end{equation}

In this paper we focus on lifts to 10D of SU(2) gauged supergravities on geometries that contain a 3-sphere. We show below that in this lifts the same isometry mixing occurs \eqref{eq:MixingLorentzGauge}, with the gauge symmetry replaced by the isometries of the internal manifold used to lift the configuration. 

Moreover, we focus on cases in which $f(x)=1$ in \eqref{eq:MetricAnsatz}. This product space structure makes the Meron configuration regular. This is to be contrasted, in the field theory contexts, with Meron configuration in 4D flat space, e.g. \cite{deAlfaro:1976qet, Drukker:2000wx}, where the size of the sphere shrinks to zero size. Similarly, in gravitational solutions containing Meron gauge fields, e.g. \cite{Canfora:2012ap, Canfora:2018ppu}, the size of the sphere shrinks at some point in space, rendering the SU(2) field strength singular. We note that the singularity in \cite{Canfora:2021nca} can be removed by a appropriate Weyl frame redefinition, yielding a regular gauge field.  In this sense, the solutions studied here avoid the usual singular behaviour associated with Meron solutions.

Before discussing two solutions explicitly, we discuss some general aspects of the 10D geometries resulting from these lifts.

\subsection{Uplifts using 3-spheres}

Let us consider a pseudo-Rimemannian manifold $(\hat{M},g)$, which
is of the form 
    \begin{align*}
        \hat{M} = M\ltimes[\text{S}^{2}\ltimes(\text{S}^{3}\times\dots\times \text{S}^{3})]
    \end{align*}
The manifold $M$ is Lorentzian and the internal manifold is defined as the semi-direct product (the metric is not block diagonal) between a 2-sphere and a set of $n$ 3-spheres. It is convenient to use the fact that $\text{S}^{3}\cong \SU{2}$ to write the 3-spheres using the left-invariant Maurer-Cartan 1-forms $\omega_{\alpha}=g_{\alpha}^{-1}\dd g_{\alpha}$ with $\alpha=1,\dots,n$. These satisfy
    \begin{equation}
        \dd \omega^{i} = -\frac{1}{2}\epsilon_{ijk} \omega^{j} \wedge \omega^{k},
    \end{equation}
for any $\alpha$. The spacetimes that we consider are of the form\footnote{Generically, the uplifted SU(2) gauge fields appear as a fiber over the 3-sphere by means of the Maurer-Cartan forms as $\omega^{i} - g A^{i}$. Since the gauge field \eqref{eq:MeronFields}, together with $\lambda = 1/2g$, what enters the 10D metric is the combination $\omega^{i} + \epsilon_{ijl}\mu^{j}d\mu^{k}$. By abuse of notation from here and below we write $A^{i} = \epsilon_{ijl}\mu^{j}d\mu^{k}$. }
    \begin{equation}\label{eq:MetricGeneral}
        \ds =\ds(M)+H_{0}\,   \ds(\text{S}^{2})+\frac{1}{4}\sum_{\alpha=1}^{n}H_{\alpha}\sum_{i=1}^{3}(\omega_{\alpha}^{i}+A^{i})^{2}
    \end{equation}
where $H_{0},H_{\alpha}$ are scalar fields on $M$. We now show how the diagonal symmetry between SO(3) spacetime and SU(2) gauge in the lower dimensional configurations \eqref{eq:MixingLorentzGauge} is realised when uplifting using 3-spheres. 

\subsubsection{Uplifted Diagonal Isometry}\label{sec:UpliftS31}

As stated before, the one forms $A^{i} = \epsilon_{ijk}\mu^{j}\dd \mu^{k}$ are dual to the Killing vectors $(J^{i})$ on the 2-sphere
    \begin{equation}
        (J^{i})^{b} = \epsilon_{ijk} \mu^{j} \partial_{a}\mu^{k} {\bar \gamma}^{a b}, 
    \end{equation}
where $a=1,2$ are coordinates along the 2-sphere and ${\bar \gamma}^{a b}$ is inverse metric of the unit-radius 2-sphere. Then, the one forms $A^{i}$ satisfy 
    \begin{equation}\label{eq:JonA}
    \mathscr{L}_{J^{i}}A_{}^{j} =-\epsilon_{ijk}A^{k}.
    \end{equation}
On the other hand, on each 3-sphere (without the fibration) there is a SU(2)$_{L}\times$SU(2)$_{R}$ isometry, generated by the left ($L^{i}$) and right ($R^{i}$) Killing vectors, which satisfy\footnote{The $L/R$ vectors are dual to the $R/L$ invariant forms.}
    \begin{equation}
        \left[ L^{i},L^{j} \right] = - \epsilon_{ijk} L^{k},\quad
        \left[ R^{i},R^{j} \right] = \epsilon_{ijk} R^{k}.
    \end{equation}
While the $R/L$ are invariant under the $R/L$ action, they are charged under the opposite one. For our purposes, we only require    \begin{equation}\label{eq:Ronw}
        \mathscr{L}_{R^{i}}\omega^{j}=\epsilon_{ijk}\, \omega^{k}.
    \end{equation}
At this point we note that this action is of the same form as \eqref{eq:JonA}, which is the key element for the existence of the  diagonal symmetry. Consider the vector
    \begin{equation}
        K^{i} =  -J^{i} + \sum_{\alpha} R_{\alpha}^{i},
    \end{equation}
where we have restored the index $\alpha$ labelling the different 3-spheres. We now compute the Lie derivative of the metric \eqref{eq:MetricGeneral} along $K^{i}$
\begin{equation}\label{eq:LieDMetric}
\begin{aligned}
\mathscr{L}_{K^{i}}\ds & =
\mathscr{L}_{K^{i}}\ds(M)
+\mathscr{L}_{K^{i}}H_{0} \,\ds(\text{S}^{2})
+H_{0}\mathscr{L}_{K^{i}}\ds(\text{S}^{2})
+\frac{1}{4}\sum_{\alpha}\mathscr{L}_{K^{i}}H_{\alpha}\sum_{i}(\omega_{\alpha}^{i}+A^{i})^{2}+ \\
 & +\frac{1}{2} \sum_{\alpha}H_{\alpha}\sum_{i}(\omega_{\alpha}^{i}+A^{i})\otimes\mathscr{L}_{K^{i}}(\omega_{\alpha}^{i}+A^{i})\,.
\end{aligned}
\end{equation}
Since the Lie derivative only has support on the 2-sphere and on the 3-spheres 
\begin{equation}
\mathscr{L}_{K^{i}}\ds(M) =\mathscr{L}_{K^{i}}H_{0}
=\mathscr{L}_{K^{i}}H_{\alpha}=0\,.
\end{equation}
On the other hand, the action of the Lie derivative is linear in the vector fields $\mathscr{L}_{X+Y}Z=\mathscr{L}_{X}Z+\mathscr{L}_{Y}Z$, we have that on the metric on the 2-sphere 
\begin{equation}
\mathscr{L}_{K^{i}}\ds(\text{S}^{2}) 
=\mathscr{L}_{-J^{i}+\sum_{\alpha} R^{i}_{\alpha}} \ds(\text{S}^{2})
=\mathscr{L}_{-J^{i}}\ds(\text{S}^{2})=0\,.
\end{equation}
Finally, the action on the fibered 3-spheres is 
\begin{equation}\label{eq:Killing1}
\mathscr{L}_{K^{j}}(\omega_{\alpha}^{i}+A^{i}) =\sum_{\beta}\mathscr{L}_{R^{j}_{(\beta)}}\omega_{\alpha}^{i}+\mathscr{L}_{-J^{j}}A^{i}
 =\epsilon_{jil}(\omega_{\alpha}^{l}+A^{l}), 
\end{equation}
from where the Lie derivative of the metric \eqref{eq:LieDMetric}
\begin{equation}
\mathscr{L}_{K^{j}}\ds  =\frac{1}{2} \sum_{\alpha}H_{\alpha}\sum_{i}(\omega_{\alpha}^{i} 
+ A^{i})\otimes\epsilon_{jil}(\omega_{\alpha}^{l}+A^{l})=0\,,
\end{equation}
Hence, $K^{i}$ is a Killing vector of the fibered metric \eqref{eq:MetricGeneral}. Let us remark that the key element for this result is that the actions of $J^{i}$ and $R^{i}$ on their dual forms is the same, \eqref{eq:JonA} and \eqref{eq:Ronw} respectively, which allows one to obtain \eqref{eq:Killing1}. Also, it is direct to check that
    \begin{equation}
        \left[ K^{i},K^{j} \right] = \epsilon_{ijk}K^{k}
    \end{equation}
so that it closes as an SU(2) algebra. The existence of these isometries, which is a diagonal combination of the SO(3) isometry of the 2-sphere and the SU(2)$_{R}$ of the 3-spheres, is the higher-dimensional realisation of the global-gauge symmetry mixing produced by the non-abelian monopole. We also highlight that it is possible to find an additional U(1) isometry, generated by
    \begin{equation}\label{eq:KillingW}
        W = \mu^{i}K^{i}.
    \end{equation}
Indeed, one can check that\footnote{We note that $W = \mu^{i}K^{i} = \mu^{i}\sum_{\alpha}R_{(\alpha)}^{i}$, since $\mu^{i}J^{i}=0$. Hence $W$ acts trivially on the 2-sphere.}
    \begin{equation}
        \mathscr{L}_{W}(\omega_{\alpha}^{i}+A^{i}) =\mathscr{L}_{W}\omega_{\alpha}^{i}
        = \mu^{j} \mathscr{L}_{\sum_{\beta}R_{(\beta)}^{j}} \omega_{\alpha}^{i}
        + \dd\mu^{j} \iota_{\sum_{\beta}R_{(\beta)}^{j}} \omega_{\alpha}^{i}
        =-\epsilon_{ijk}\mu^{j}(\omega_{\alpha}^{k}+A^{k}),
    \end{equation}
from where is direct to check that $W$ is indeed an isometry of the manifold and that it commutes with all other generators. With this, the complete set of isometries of the spacetime \eqref{eq:MetricGeneral} is
    \begin{equation}\label{eq:IsomM}
        \mathrm{Isom}(M)\times\underbrace{\mathrm{SU}(2)\times\dots\times \mathrm{SU}(2)}_{n-\mathrm{times}}\times \mathrm{SU}(2)_{D} \times \text{U(1)}_{W}
    \end{equation}
where, the $n$ copies of SU(2) factors correspond to the SU(2)$_{L}$ of each 3-sphere which is not broken by the fibrations (as $\omega^{i}$ invariant under $L^{i}$). It is also possible to show that this is the case by following Appendix C of \cite{Legramandi:2020txf}.

When lifting to Type II, we need to ensure that the dilaton and the NS-NS and RR fluxes are also invariant under the \eqref{eq:IsomM} for it to be a symmetry of the full configuration. For the dilaton we just require that it does not depend on the 2- and 3-spheres. In order to construct $p$-forms which are invariant under $\SU{2} \times ... \times \SU{2} \times \text{SU}(2)_{D}$ we can use the invariant tensors $\delta_{ij}$ and $\epsilon_{ijk}$ to construct  $\SU{2} \times ... \times \SU{2} \times \text{SU}(2)_{D}$ scalars using $(\mu^{i},d\mu^{j}, \omega^{k})$. Such $p$-forms must be the only ones appearing in the 10D fluxes.

\subsubsection{A Possible Generalization}\label{sec:Uplifts32}

We can also consider a Riemannian manifold $\mathscr{B}$ such that it admits a projection map $\pi : \mathscr{B} \to \text{S}^{2}$ and there is a set of Killing vectors $\hat{J}^{i}$ realizing the $\text{SO}(3)\subset\mathrm{Isom}(\mathscr{B})$. More precisely, $\hat{J}^{i}$ is the lift of the 2-sphere Killing vectors to $\mathscr{B}$ (it can possibly contain terms along directions of $\mathscr{B}$ which are not of the 2-sphere). Importantly, since $A^{i}$ is only takes values on the 2-sphere co-tangent space
\begin{align}
\mathscr{L}_{\hat{J}^{i}}A^{j} & =\epsilon_{ijk}A^{k},
\end{align}
as before. When this happens, the proof on the invariance of the metric under the diagonal symmetry is a straightforward generalization of the previous one. The metric under consideration now takes the form
\begin{align}
\ds(\hat{M}) & =\ds(M)+H_{0}\, \ds(\mathscr{B}) + \sum_{\alpha}H_{\alpha}\sum_{i}(\omega_{\alpha}^{i}+A^{i})^{2}\,,
\end{align}
and its invariance under $K^{i}=\sum_{\alpha}R_{\alpha}^{i} - \hat{J}^{i}$ is direct. 

We now focus on solutions of the form described in this section, showing one example for each of the two constructions discussed above.  

\section{I-branes on \texorpdfstring{$\Sp{2}$}{S2} with a Meron}\label{sec:Ibranes}

The first family of configuration, that belongs to class of backgrounds discuss in Section \ref{sec:UpliftS31}, is obtained by lifting the double-Meron solution of the 4D SU(2)$\times$SU(2) gauged supergravity \cite{Canfora:2021nca} to Type IIB. It corresponds to a holographic dual of the I-brane theory \cite{Itzhaki:2005tu} on $\Sp{2}$ with a magnetic monopole on the 2-sphere. In string frame, the metric is the direct product between a 2-dimensional manifold with coordinates $(t,\rho)$, and an internal 8-dimensional manifold
\begin{equation}
    \dd s^{2}  =-(1-me^{2\Phi})\dd t^{2}+\frac{\dd\rho^{2}}{1-me^{2\Phi}}+\dd s^{2}(M_{8})\,,
\end{equation}
where the  $8$-dimensional manifold is of the form $\Sp{2}\ltimes(\Sp{3}\times \Sp{3})$ given by 
    \begin{align}
        \dd s^{2}(M_{8}) & =\ell_{\Sp{2}}^{2}\, \dd\mu^{i}\dd\mu^{i}+\frac{N_{B}}{4}h_{A}^{i}h_{A}^{i}+\frac{N_{A}}{4}h_{B}^{i}h_{B}^{i}\,, \label{8D manifold} \\
        h_{\alpha}^{i} & =\omega_{\alpha}^{i}+\epsilon_{ijk}\mu^{j}\dd\mu^{k}\,,\qquad
         \mu^i \mu^i = 1 \, , \qquad \alpha=A,B\,.
    \end{align}
and the matter fields supporting the configuration are 
\begin{align}
H_{3} & =\frac{N_{B}}{4}[\mu^{i}h_{A}^{i} \wedge \vol(\Sp{2})+\frac{1}{3!}\epsilon_{ijk}h_{A}^{ijk} ] + \frac{N_{A}}{4}[\mu^{i}h_{B}^{i}\wedge \vol(\Sp{2})+\frac{1}{3!}\epsilon_{ijk}h_{B}^{ijk} ]\,,\\
\Phi & =-\mathfrak{C}\, \rho\,.
\end{align}
The dilaton slope and the $\Sp{2}$ radius are 
\begin{align}
 & \mathfrak{C}^{2}=\frac{1}{N_{A}}+\frac{1}{N_{B}}+\frac{1}{N_{A}+N_{B}}\,,\qquad
 \ell_{\Sp{2}}^{2}=\frac{N_{A}}{4}+\frac{N_{B}}{4}\,.
\end{align}

The configuration is characterized by the integration constants $m,N_{A},N_{B}$. The constants $N_{A},N_{B}$ are flux of the 3-form on each 3-sphere. 
    \begin{equation}
        \frac{1}{(2\pi)^{2}}\int_{\Sp{3}_{A}} H_{3}= N_{B} \,, \qquad \frac{1}{(2\pi)^{2}}\int_{\Sp{3}_{B}} H_{3}= N_{A}
    \end{equation}
Notice that the configuration is symmetric under $A\leftrightarrow B$. The 2-dimensional manifold with coordinates $(t,\rho)$ is asymptotically flat, as $\rho\to+\infty$. For $m>0$ it corresponds to a black hole with the horizon located at $\rho_{H}=\log(m)/2\mathfrak{C}$. Upon compactification, the 2-dimensional manifold corresponds to a 2-dimensional black hole (similar configuration were studied in \cite{Mann:1991ny,Lemos:1994fn}), and corresponds to the coset SL(2,$\mathbb{R}$)$_{k}$/U(1), with level $k=\mathfrak{C}^{-2}$ \cite{Witten:1991yr}. The black hole temperature and the Bekenstein-Hawking entropy are\footnote{These quantities are to be computed in Einstein frame}
\begin{align}
T & =\frac{\mathfrak{C}}{2\pi}\,,\qquad S=4\pi^{3}mN_{A}^{3/2}N_{B}^{3/2}\ell_{\Sp{2}}^{2}\,.
\end{align}
From these quantities, one can show that the following relation between variation of the parameters holds
\begin{align}
\delta M & =T\delta S+\Omega_{A}\delta N_{A}+\Omega_{B}\delta N_{B}\,,
\end{align}
with
\begin{equation}
M=2\pi^{2}mN_{A}^{3/2}N_{B}^{3/2}\ell_{S^{2}}^{2}\mathfrak{C}\,, \quad
\Omega_{A} =\frac{S}{2\pi}\frac{\partial\mathfrak{C}}{\partial N_{A}}\,,\quad
\Omega_{B} =\frac{S}{2\pi}\frac{\partial\mathfrak{C}}{\partial N_{B}}\,. 
\end{equation}
The finite thermodynamic quantities that we have derived satisfy a Smarr formula $M=TS$, which implies that the free energy vanishes identically. This is a consequence of the fact that the black hole temperature is independent on the mass parameter $m$.

The 8-dimensional manifold $M_8$ defined in \eqref{8D manifold} is smooth and free of curvature singularities. The 1-forms $h^i_\alpha (\alpha =A,B)$ define globally well-defined SU(2) fibrations over the $\text{S}^2$, with $\omega^i_\alpha$ the left-invariant Maurer-Cartan forms on SU(2). Since the coefficients of the bilinear $h^i_A h^i_A$ and $h^i_B h^i_B$ in the metric are strictly positive constants and independent of the base coordinates, the fibres do not degenerate anywhere on $\text{S}^2$. As a consequence, the Riemannian manifold $M_8$ is regular and free of conical singularities. Its Ricci tensor is given by
\begin{align}
R(M_{8})_{\overline{\mu\nu}}\dd x^{\overline{\mu}}\otimes\dd x^{\overline{\nu}} & =\frac{1}{2}(\dd\mu^{i}\dd\mu^{i}+h_{A}^{i}h_{A}^{i}+h_{B}^{i}h_{B}^{i})+\frac{8}{(N_{A}+N_{B})^{2}}\mathsf{A}\otimes\mathsf{A}
\end{align}
where $\overline{\mu}$ labels coordinates of the 8D manifold. The 1-form $\mathsf{A}$, given by
\begin{align}
\mathsf{A} & =\frac{1}{4}(N_{B}\mu^{i}\omega_{A}^{i}+N_{A}\mu^{i}\omega_{B}^{i})\,,
\end{align}
is the dual of generator of the additional U(1)$_{W}$ isometry found in \eqref{eq:KillingW}. This shows that $M_8$ is not an Einstein manifold. The trace of the Ricci tensor is $6\mathfrak{C}^{2}$.

The symmetry of the configuration, that in general is a subgroup of the isometry group, is given by
\begin{equation}\label{symmetry double Meron}
\mathbb{R}\times {\rm SU(2)}_{L}\times {\rm SU(2)}_{L}\times \text{SU}(2)_{D} \times \text{U(1)}_{W}
\end{equation}
where $\mathbb{R}$ is associated to time translations, and SU(2)$_D$ is generated by a combination of  SU(2)$_{\text{R}}$ Killing vectors of the 3-spheres ($R^{i}_{\alpha}$) and the SO(3) ones of the  $\Sp{2}$ ($J^{i}$), as was explained in Section \ref{sec:UpliftS31}.

The ten-dimensional space develops a curvature singularity for $m \neq 0$ located at $\rho \to -\infty$, as indicated by the Kretschmann scalar:
\begin{equation}
    R_{\mu \nu \rho \sigma } R^{\mu \nu \rho \sigma} = \frac{12 \left(2 N_A^3 N_B+11 N_A^2 N_B^2+2 N_A N_B^3+N_A^4+N_B^4\right)}{N_A^2 N_B^2 \left(N_A+N_B\right){}^2} + 16 \mathfrak{C}^4 m^2 e^{-4 \mathfrak{C} \rho}
\end{equation}
In the same region the scalar $e^{\Phi}$ is divergent. However, this singularity is hidden behind the event horizon, and the degree of divergence is controlled by the parameter $m$. 

In the limit $m=0$, the ten-dimensional manifold is regular for $\rho \in \mathbb{R}$, since the geometry reduces to the direct product $\mathbb{R}^{1,1} \times M_8$. Nevertheless, the dilaton still diverges as $\rho \to - \infty$, making the background singular from the string perspective. This however, is the usual case in backgrounds constructed purely using NS-branes. In the next section we show that this limit preserves supersymmetry.


\subsection{Supersymmetric Solution}

The configuration is supersymmetric for $m=0$. In this case, it might seem that the isometry group of the 2-dimensional manifold get enhanced to Poincare, but the explicit dependence of the dilaton in one of the coordinates breaks it, leading to the symmetry group given in (\ref{symmetry double Meron}). 

To integrate the Killing spinors, we start with the natural set of vielbeins
    \begin{align}
        e^{0} & =\dd t\,,\qquad 
        e^{1}=\dd\rho\,,\qquad e^{2}=\ell_{\Sp{2}}\dd\theta\,,\qquad    e^{3}=\ell_{\Sp{2}}\sin\theta\dd\varphi\,,\label{natural vielbein}\\
        e^{i+3} & =V_{A}^{i}\equiv\frac{\sqrt{N_{B}}}{2}h_{A}^{i}\,,\qquad 
        e^{i+6}=V_{B}^{i}\equiv\frac{\sqrt{N_{A}}}  {2}h_{B}^{i}\,,\qquad i=1,2,3.\nonumber 
    \end{align}
It is convenient to perform a local rotation $\mathbf{R}\in\mathrm{SO(6)}\subset\mathrm{O}(1,9)$ on the vielbeins (\ref{natural vielbein}) as follows
    \begin{align}
        \tilde{e}^{a}=\mathbf{R}^{a}{}_{b}e^{b}\,,\qquad
        (\mathbf{R}^{a}{}_{b}) & =\left(
        \begin{array}{c|cc|cc}
            \mathds{1}_{4} &  & \\
            \hline  & \mathds{1}_{2} & \\
            &  & S_{A} &  & -S_{B}\\
            \hline  &  &  & \mathds{1}_{2}\\
            &  & S_{B} &  & S_{A}
        \end{array}
        \right)\left(
        \begin{array}{c|c|c}
        \mathds{1}_{4} & \\
            \hline  & \mathsf{R}\\
            \hline  &  & \mathsf{R}
            \end{array}\right),
        \end{align}
where $\mathsf{R}\in\mathrm{SO}(3)$ is defined in term of $\mu^{i}$ and its derivatives, $S_{A},S_{B}$ are constants. They are given by
    \begin{align}
        \mathsf{R} & =\left(\begin{array}{c}
        \partial_{\theta}\mu^{i}\\
        \frac{1}{\sin\theta}\partial_{\varphi}\mu^{i}\\
        \mu^{i}
    \end{array}\right)\,,\qquad 
    S_{A}=\sqrt{\frac{N_{A}}{N_{A}+N_{B}}}\,,\qquad S_{B}=\sqrt{\frac{N_{B}}{N_{A}+N_{B}}}\,.
\end{align}
We consider the basis ${\tilde e}^a$, and therefore, define the Clifford map as ${\tilde e}^a \mapsto \Gamma^a$. In Type II in the pure NS sector, the supersymmetry conditions reduce to two independent system of equations for two decoupled Majorana-Weyl spinors $(\epsilon^{1},\epsilon^{2})$ of opposite (IIA) or same (IIB) chiralities. We take $\epsilon^{1}$ to be of positive chirality, while $\epsilon^{2}$ is of negative/positive chirality in Type II A/B. The supersymmetry variations are given by
    \begin{equation}
    \begin{aligned}
        \delta \lambda^{1} &= \left( \dd \Phi - \frac{1}{2}H_{3}\right)_{/} \epsilon^{1}, \quad 
        \delta \Psi^{1}_{\mu} = \left( D_{\mu} - \frac{1}{4}H_{\mu}\right)_{/}\epsilon^{1}, \label{eqs to solve eps1}\\
        \delta \lambda^{2} &= \left( \dd \Phi + \frac{1}{2}H_{3}\right)_{/} \epsilon^{2}, \quad 
        \delta \Psi^{2}_{\mu} = \left( D_{\mu} + \frac{1}{4}H_{\mu}\right)_{/}\epsilon^{2}
    \end{aligned}
    \end{equation}
where $D_{\mu} = \partial_{\mu} + \frac{1}{4}\omega_{\mu}^{ab}\Gamma_{ab}$ is the spinorial covariant derivative. It is possible to obtain algebraic integrability conditions from the gravitino variation. First, we start by defining the 1-form connections
    \begin{equation}
        W^{1} = \frac{1}{4}\left(\omega_{\mu} - H_{\mu} \right)_{/}\dd x^{\mu},\qquad
        W^{2} = \frac{1}{4}\left(\omega_{\mu} + H_{\mu} \right)_{/} \dd x^{\mu} \,,
    \end{equation}
which have
    \begin{equation}
        \Xi^{I} = dW^{I} + W^{I}\wedge W^{I}, \qquad I=1,2,
    \end{equation}
as field strengths. With this, the gravitino variations can be written as
    \begin{equation}
        \mathcal{D}^{(1)}_{\mu} \epsilon^{1}=0, \quad
        \mathcal{D}^{(2)}_{\mu} \epsilon^{2}=0, \quad 
        \text{with}\quad \mathcal{D}^{(I)}_{\mu} = \partial_{\mu} + W^{I}_{\mu}. 
    \end{equation}
We now consider the commutator of two covariant derivatives
    \begin{equation}
        \left[ \mathcal{D}^{1}_{\mu},\mathcal{D}^{1}_{\nu} \right]\epsilon^{1} \equiv  \Xi^{1}_{\mu\nu} \epsilon^{1},\quad
        \left[ \mathcal{D}^{2}_{\mu},\mathcal{D}^{2}_{\nu} \right]\epsilon^{2} \equiv  \Xi^{2}_{\mu\nu} \epsilon^{2}
    \end{equation}
since the left hand side of each of these equations is zero, then each of the components of $\Xi^{I}$ (recall that they are matrices due to the Clifford map) acting on the spinor defines an algebraic condition on the spinor
    \begin{equation}
         \Xi^{1}_{\mu\nu} \epsilon^{1} = 0, \quad
          \Xi^{2}_{\mu\nu} \epsilon^{2}=0.
    \end{equation}
In order to find non-trivial solutions, each of these components must have zero determinant. We find that $\det(\Xi^{2}_{34} )\neq 0$, which sets $\epsilon^{2}=0$. On the other hand, all the components of $\Xi^{1}$ have zero determinant. In fact, all but one of its components are zero. That component leads to the following condition on $\epsilon^{1}$
\begin{align*}
(1-\Gamma^{4578})\epsilon^{1} & =0\,.
\end{align*}
The solution is
\begin{align}
\epsilon^{1} & =P_{1}\eta\,, \hspace{1.5cm}P_{1} \equiv\frac{1}{2}(1+\Gamma^{4578}) \label{sol dW+WW} \, .
\end{align}
We now turn to the dilatino variation, upon using the solution (\ref{sol dW+WW}),
it can be recasted as
\begin{align}
-\frac{1}{\mathfrak{C}}\Gamma^{1}\delta\lambda^{1} & =\left(1-\frac{1}{\sqrt{N_{A}+N_{B}}\mathfrak{C}}\Gamma^{1239}-\frac{1}{\mathfrak{C}}\sqrt{\frac{1}{N_{A}}+\frac{1}{N_{B}}}\Gamma^{1456}\right)\epsilon^{1}=0
\end{align}
This is again a projector acting on the spinor $\epsilon^{1}$ whose
solution is
\begin{align}
\epsilon^{1} & =P_{1}P_{2}\eta\,,\\
P_{2} & \equiv\frac{1}{2}\left(1+\frac{1}{\sqrt{N_{A}+N_{B}}\mathfrak{C}}\Gamma^{1239}+\frac{1}{\mathfrak{C}}\sqrt{\frac{1}{N_{A}}+\frac{1}{N_{B}}}\Gamma^{1456}\right)
\end{align}
The two projectors commute $[P_{1},P_{2}]=0$, and they do not depend on coordinates which simplifies the integration of the differential equations. We now proceed to the Killing spinor equation for $\epsilon^{1}$ \eqref{eqs to solve eps1}, which comes from the gravitino variation. Upon using the projectors, it takes the form
\begin{align}
\left\{ \dd+\frac{1}{\sqrt{N_{A}+N_{B}}}\left[\Gamma^{39} \tilde{e}^{2}-(\Gamma^{29}+\Gamma^{23}\cot\theta_{1}) \tilde{e}^{3}\right]\right\} \epsilon^{1} & =0\,.
\end{align}
The only non-trivial equations are
\begin{align}
\partial_{\theta}\epsilon^{1}+\frac{1}{2}\Gamma^{39}\epsilon^{1} & =0\,,\\
\partial_{\phi}\epsilon^{1}-\frac{1}{2}\Gamma^{23}(\cos\theta+\Gamma^{39}\sin\theta)\epsilon^{1} & =0\,.
\end{align}
Which can be directly integrated to find Killings spinor to be
\begin{align}
\epsilon^{1} & =e^{-\frac{\theta}{2}\Gamma^{39}}e^{\frac{\phi}{2}\Gamma^{23}}P_{1}P_{2}\eta_{0}\,,
\end{align}
where $\eta_{0}$ is a constant spinor Majorana-Weyl with 16 real components. Since each projector halves the number of independent components the background preserves 4 supercharges. 

This spinor is uncharged under the $\text{SU(2)}_{L_{A}} \times \text{SU(2)}_{L_{B}}$ and U(1)$_{W}$ symmetries 
    \begin{equation}
        \mathcal{L}_{L^{i}_{A,B}} \epsilon^{1} = 0, \qquad
        \mathcal{L}_{W}\epsilon^{1} = 0.
    \end{equation}
On the other hand, writing the four independent components of the spinors as
    \begin{equation}
        \epsilon^{1} = \sum^{4}_{a} \chi_{a},
    \end{equation}
it is possible to check that the spinor is charged under the diagonal SU(2)$_{D}$ symmetry 
    \begin{equation}
        \mathcal{L}_{K^{i}} \chi^{a} = -\left( \Sigma_{i}\right)^{a}_{\phantom{a}b} \chi^{b},
    \end{equation}
where the matrices $\Sigma_{i}$ are
    \begin{equation}
        \Sigma_{1} = \begin{pmatrix}
            0 & 0 & 0 & \frac{1}{2} \\
            0 & 0 & \frac{1}{2} & 0 \\
            0 & -\frac{1}{2} & 0 & 0 \\
            -\frac{1}{2} & 0 & 0 & 0 
        \end{pmatrix}, \quad
        \Sigma_{2} = \begin{pmatrix}
            0 & 0 & \frac{1}{2} & 0 \\
            0 & 0 & 0 & -\frac{1}{2} \\
            -\frac{1}{2} & 0 & 0 & 0 \\
            0 & \frac{1}{2} & 0 & 0 
        \end{pmatrix}
        \Sigma_{3} = \begin{pmatrix}
            0 & \frac{1}{2} & 0 & 0 \\
            -\frac{1}{2} & 0 & 0 & 0 \\
            0 & 0 & 0 & \frac{1}{2} \\
            0 & 0 & -\frac{1}{2} & 0 
        \end{pmatrix}.
    \end{equation}
These closed the $\mathfrak{su}(2)$ algebra
    \begin{equation}
        \left[ \Sigma_{i},\Sigma_{j} \right] = \epsilon_{ijk} \Sigma_{k} \, .
    \end{equation}
From here, there is a similarity transformation that maps these matrices to the ones in \cite{Lozano:2019emq}, which explicitly  shows that the Killing spinors $\chi^{a}$ transform in the $2+\bar{2}$ of SU(2)$_{D}$.

\section{\texorpdfstring{$\AdS{3}\times\Sp{2}$}{AdS3} with a Meron}\label{sec:Section3}

In this section we derive a new solution of a 5D $\mathcal{N}=4$ SU(2)$\times$U(1) gauged supergravity \cite{Romans:1985ps}, which contains a Meron configuration for the SU(2) gauge field. We then proceed to lift this solution to Type IIB using \cite{Lu:1999bw}. This lift fits in the class of embeddings discusses in Section \ref{sec:Uplifts32}, and we show explicitly the construction of the diagonal symmetry generators. 

\subsection{5D SU(2)\texorpdfstring{$\times$}{x}U(1) Gauged Supergravity}\label{sec:GaugedSugra}

The bosonic sector of the theory consists of (we follow the notation of \cite{Lu:1999bw}): the metric tensor, two 2-forms $A^{\alpha}_{\phantom{\alpha}\mu\nu}$, a SU(2) gauge field $A^{a}_{\mu}$ with gauge coupling $g_{2}$, a U(1) gauge field $B_{\mu}$ with gauge coupling $g_{1}$, and a scalar $\phi$. By setting the 2-forms $A^{\alpha}$ to zero, and focusing on the $g_{1} = g$ and $g_{2} = \sqrt{2}g$ sector, the bosonic action is
    \begin{equation}
    \begin{aligned}
        \mathcal{L} &=   R\star1-3X^{-2}\dd X\wedge\star\dd X-\frac{1}{2}X^{4}G_{(2)}\wedge\star G_{(2)} - \frac{1}{2} X^{-2}F_{(2)}^{i}\wedge\star F_{(2)}^{i}\\
        &\phantom{=} -\frac{1}{2}F_{(2)}^{i}\wedge F_{(2)}^{i}\wedge B_{(1)} +4g(gX^{2}+gX^{-1})\star1\,,
    \end{aligned}
    \end{equation}
where $X=e^{-\frac{\phi}{\sqrt{6}}}$, $G_{(2)} = \dd B_{(1)}$ and the non-Abelian field strength is given by
    \begin{equation}
        F_{(2)}^{i}  =\dd A^{i}+\frac{\sqrt{2}}{2}g\varepsilon_{ijk}A^{j}\wedge A^{k}.
    \end{equation}
The convention for the SU(2) generators is the same as in Section \ref{sec:Section2},  $T^{i} = -\frac{i}{2}\sigma^{i}$, with structure constants to $f_{ijk}=\epsilon_{ijk}$. 

We now propose an ansatz for the spacetime. In order to have the Meron configuration, it has to contain a 2-sphere. The three remaining directions must span a space of negative curvature. With these requirements, the simplest possible ansatz for the background fields is
    \begin{subequations}
    \begin{align}
        \dd s^{2}= & \ellAdS^{2} \ds(\text{AdS}_{3}) + \ellS^{2}(\dd\theta^{2}+\sin^{2}\theta\dd\varphi^{2})\,,\\
        A_{(1)}= & \lambda\,  U^{-1}\dd U ,\\
        B_{(1)}= & q_{m}\cos\theta\dd\varphi\,,\\
        \phi= & 0\, ,
    \end{align}
    \end{subequations}
where $\lambda$ and $q_{m}$ are constant and the $\ds(\text{AdS}_{3})$ is of unit radius. Besides the SU(2) Meron gauge field, we have turned on the 1-form $B_{(1)}$ as a monopole on the 2-sphere. This is useful since it is what allows one to set the scalar $\phi$ to the minimum of the potential. More precisely, recall from Section \ref{sec:Section2} that $F_{(2)}^{i}=-\frac{1}{\sqrt{2}g}\mu^{i}\Vol(\text{S}^{2})$ (with $g_{2}=\sqrt{2}g$). It is direct to show that\footnote{At this point one may wonder if the Meron configuration can be gauge transformed to an Abelian monopole configuration $A^{1}=A^{2}=0$, $A^{3}\sim \cos(\theta)d\varphi$. This gauge transformation is singular and therefore changes the properties of the system.}
    \begin{equation}\label{eq:FieldStrengths}
        F_{(2)}^{i}\wedge\star F_{(2)}^{i} \sim \frac{1}{2} G_{(2)}\wedge\star G_{(2)},
    \end{equation}
where the proportionality constant depends on $\lambda$ and $q_{m}$. We now turn to the scalar equation of motion
    \begin{equation}
        \dd(X^{-1}\star\dd X)-\frac{1}{3}X^{4}G_{(2)}\wedge\star G_{(2)}+\frac{1}{6}X^{-2}(F_{(2)}^{i}\wedge\star F_{(2)}^{i}+\AA_{(2)}^{\alpha}\wedge\star\AA_{(2)}^{\alpha}) +\frac{4}{3}g^{2}(X^{2}-X^{-1})\star1 =0
    \end{equation}
we see that exact combination of \eqref{eq:FieldStrengths} appears, allowing one to set the scalar to the minimum, provided the constants $\lambda$ and $q_{m}$ are related. The equations of motion fix the constants as
    \begin{equation}\label{eq:Constants5D}
        q_{m} =  \frac{1}{2g}\,, \quad
        \lambda = \frac{1}{2\sqrt{2}\, g}\,,\quad
        \ellAdS^{2} = \frac{1}{g^{2}}\frac{2}{7+\sqrt{5}}, \quad
        \ellS^{2} = \frac{1}{g^{2}}\frac{1}{2(1+\sqrt{5})} \,,
    \end{equation}
leaving no free parameters.
\subsection{Type IIB Uplift}

In order to embed this solution in 10D, we follow  \cite{Lu:1999bw}. The 5D solution is lifted on a 5-sphere parametrisation that makes the SU(2)$\times$U(1) symmetry manifest. The 10D background contains only the metric and 5-form and reads, 
    \begin{equation}
    \begin{aligned}
        \ds_{10\text{D}} &= \ds_{5\text{D}} 
            +  \frac{1}{g^{2}}\left[ \dd \xi^{2} 
            + \frac{1}{4} \sin^{2}\xi  \left(\dd\tau + \cos(\theta)\dd\varphi\right)^{2}  
            + \frac{1}{4}\cos^{2}\xi \sum^{3}_{i=1} \left(\omega^{i} -\sqrt{2}g A^{i}\right)^{2}   \right],\\
        F_{5} &= (1+\star_{10})G_{5},
    \end{aligned}
    \end{equation}
where
    \begin{align}
        G_{5} &= 4g  \Vol(\mathcal{M}_{5D}) 
            + \frac{\cos^{2}(\xi)}{8\sqrt{2} g^{2}}\star_{5} F^{i}\wedge h^{j}\wedge h^{k} \epsilon_{ijk}
            - \frac{\sin(\xi)\cos(\xi)}{2\sqrt{2}g^{2}} \star_{5} F^{i} \wedge h^{i} \wedge d\xi \\
            &\quad -\frac{\sin\xi\cos\xi}{g^{2}}\star_{5}\dd B\wedge\dd\xi\wedge(\dd\tau-gB) \, . \notag
    \end{align}
for $h^{i} = \omega^{i} - \sqrt{2} g A^{i}$. The internal manifold with coordinates $(\tau,\xi,\psi,\theta_{2},\phi_{2})$ corresponds to a 5-sphere which is written as the product $[0,\pi/2] \ltimes (\Sp{1}\times \Sp{3})$, where the interval is parameterized by $\xi$ and the $\Sp{1}$ by $\tau$, with $\tau\sim \tau +4\pi$. The $\Sp{3}$ is written in the Hopf fibration form in terms of the SU(2) left-invariant Maurer-Cartan
forms
\begin{equation}
\begin{aligned}
    &\omega_1=\cos\psi \dd\theta_2+\sin\theta_2\sin\psi \dd\phi_2, \\
    &\omega_2=-\sin\psi \dd\theta_2 +\sin\theta_2\cos\psi \dd\phi_2, \\ 
    &\omega_3=\dd\psi +\cos\theta_2 \dd\phi_2,
\end{aligned}
\end{equation}

There is a flux of the 5-form along the 5-sphere (we work in units such that $\alpha' = g_{s}=1$)
    \begin{equation}
        N = \frac{1}{(2\pi)^{4}}\int_{\Sp{5}} F_{5} = \frac{1}{4\pi g^{4}}.
    \end{equation}
    
Explicitly, one can find the 4-form potential for the background
\begin{align}
C_{4} & = g^{-1} \Vol(\mathrm{AdS}^{3})\wedge\left[\frac{2}{1+\sqrt{5}}\cos\theta\dd\varphi+(1+\sqrt{5})\frac{\cos^{2}\xi}{4}(\dd\tau+\cos\theta\dd\varphi+\mu^{i}\omega^{i})\right]-\\
 &\quad -\frac{\cos^{3}\xi}{16g^4}[4\sin\xi\,\dd\xi\wedge( \dd\tau+\cos\theta\dd\varphi)+\cos\xi\,\Vol(S^{2})]\wedge(\epsilon_{ijk}\mu^{i}\dd\mu^{j}\wedge\omega^{k}-\cos\theta_{2}\dd\varphi_{2}\wedge\dd\psi)-\nonumber \\
 &\quad -\frac{\cos^{2}\xi\sin^{2}\xi}{8g^4}\mu^{i}\omega^{i}\wedge\Vol(S^{2})\wedge\dd\tau\nonumber 
\end{align}

Interestingly, the direct product structure of the 5-dimensional manifold of Section \ref{sec:GaugedSugra} allows one to write the 10-dimensional metric as a direct product between AdS$_3$ and a compact 7-dimensional manifold M$_7$
    \begin{equation}
        \dd s^{2}_{10D} = \ellAdS^{2} \dd s^{2} (\AdS{3}) + \frac{1}{g^2} \dd s^{2} (\text{M}_{7}),
    \end{equation}
where
    \begin{equation}
    \begin{aligned}
        \dd s^{2}(\text{M}_{7}) &=  \dd \xi^{2} + 
        \Lambda^{2}(\dd\theta^{2}+\sin^{2}\theta\dd\varphi^{2})
            + \frac14 \sin^{2} \xi \left(\dd\tau + \cos(\theta)\dd\varphi\right)^{2}  \\
            &\phantom{=}
            + \frac14\cos^{2} \xi \sum^{3}_{i=1} \left(\omega^{i} +\epsilon_{ijk}\mu^j \dd \mu^k \right)^{2} \,,
    \end{aligned}   
    \end{equation}
with $\Lambda^{2}=\frac{1}{2(1+\sqrt{5})}$. The range of the coordinates are $\xi\in [0,\pi/2],\, \theta\in [0,\pi],\, \varphi\sim \varphi + 2 \pi, \, \tau \sim \tau +4 \pi $ and for the 3-sphere are $\theta_2 \in [0,\pi], \, \varphi_2 \sim \varphi_2 + 2\pi \,, \, \psi \sim \psi + 4 \pi$. 

The 3-dimensional submanifold obtained by fixing $(\xi,\theta_1,\varphi_2,\psi)$ to constants is a squashed 3-sphere written in the Hopf fibration coordinates $(\theta,\phi,\tau)$, where the base manifold is the 2-sphere and the fiber is the $\Sp{1}$. This space corresponds to $\mathscr{B}$ in the notation of Section \ref{sec:Uplifts32}. The Killing vectors on the 2-sphere $J^{i}$ are lifted to Killing vectors on the squashed 3-sphere as
    \begin{equation}\label{eq:defhatJ}
        \hat{J}^{i} = (J^{i})^{a} \partial_a - 2g B_{b} {\bar\gamma}^{b a}\partial_{\tau}.
    \end{equation}
with $B_{a}$ the components of $B_{(1)}$ along the 2-sphere. On the other hand, on the 3-sphere $(\theta_{2},\varphi_{2},\psi)$ without fibrations, there are two sets of Killing vectors $L^{i}$ and $R^{i}$ of the SU(2)$_{R}\times$ SU(2)$_{L}$ symmetry. As explained in Section \ref{sec:Section2}, the fibration by $\epsilon_{ijk}\mu^{j}\dd \mu^{k}$ mixes the SU(2)$_{R}$ with the isometries of the 2-sphere, while preserving the SU(2)$_{L}$. The Killing vector of the diagonal symmetry is not obtained by using \eqref{eq:defhatJ}, but rather
    \begin{equation}
        \tilde{L}^{i} = \hat{J}^{i} - \mu^{i}\partial_{\tau}.
    \end{equation}
In terms of these, the diagonal symmetry SU(2)$_{D}$ is generated by
    \begin{equation}
        K^{i} = R^{i} + \tilde{L}^{i},  \quad 
        \left[ K^{i},K^{j} \right] = \epsilon_{ijk} K^{k}.
    \end{equation}

The vectors $\tilde{L}^{i}$ are actually related to the generators of the SU(2)$_{L}$ symmetry on the squashed 3-sphere $(\theta,\varphi,\tau)$. This can be understand as follows: the 3D metric on this sphere be can obtained by considering the group element $g=e^{\varphi T_3}e^{\theta T_1} e^{\tau T_3}\in {\rm SU}(2)$ and their left-invariant 1-forms, $\hat{\omega}=g^{-1} \dd g$. Then, the squashed metric is obtained by
    \begin{equation}
        \ds_{3D} = \gamma_{ij}\hat{\omega}^{i}\hat{\omega}^{j}, \quad
        \gamma_{ij} = \text{diag}\left( \Lambda^{2},\Lambda^{2},\frac{1}{4}\sin^{2}(\xi) \right).
    \end{equation}
Since we are using left-invariant forms to construct the metric, the SU(2)$_{L}$ symmetry, generated by $\hat{L}^{i}$ is preserved. On the other hand, since $\gamma_{ij}$ is not invariant under SU(2)$_{R}$, this symmetry is broken to U(1), generated by $\partial_{\tau}$. The relation between $\tilde{L}^{i}$ and $\hat{L}^{i}$ is as follows 
    \begin{equation}
        \widetilde{L}^i = S^i{}_j \hat{L}^{j} \, \qquad S = \left(\begin{array}{ccc}
            0 & 1 & 0\\
            -1 & 0 & 0\\
            0 & 0 & -1
        \end{array}\right) \in {\rm O}(3).
    \end{equation}
We find that $W=\mu^{i}K^{i}$ is also a Killing vector that generates a U(1) symmetry. With these considerations, the isometries of M$_{7}$ is
    \begin{equation}\label{eq:IsomM7}
        \text{SU(2)}_{L} \times \text{SU(2)}_{D} \times \text{U(1)} \times \text{U(1)}_{W}.
    \end{equation}
We note that $F_{5}$ is also invariant under these transformations, so they are a true symmetry of the entire background.

Finally, we note that $M_7$ is not Einstein. Indeed, considering the following orthonormal basis
\begin{align}
\eta^{1} & =\dd\xi\,,\qquad
\eta^{2}=\Lambda\dd\theta\,,\qquad
\eta^{3}=\Lambda \sin\theta\dd\varphi \,,\qquad
\eta^{4}=\frac{\sin\xi}{2}(\dd\tau+\cos\theta\dd\varphi)\nonumber \\
\eta^{i} & =\frac{1}{2}\cos\xi(\omega^{i}+\epsilon_{ijk}\mu^{j}\dd\mu^{k})\,,\qquad i=5,6,7\,,
\end{align}
the Ricci tensor of M$_7$ can be written as
\begin{align*}
R(\text{M}_{7})_{IJ}\eta^{I}\otimes\eta^{J} & =4\eta^{I}\otimes\eta^{I}+(3+\sqrt{5})\mathsf{A}\otimes\mathsf{A}+(5-\sqrt{5})(\eta^{2}\otimes\eta^{2}+\eta^{3}\otimes\eta^{3})
\end{align*}
where $I,J=1,\dots, 7$. The last factor is showing that the curvature of the $\text{S}^2$ is different from the $\text{S}^5$ by a numerical factor. The 1-form $\mathsf{A}$ of constant norm is given by 
\begin{align}
\mathsf{A} & =\frac{1}{2}\sin^{2}\xi(2\dd\tau+\cos\theta\dd\varphi)-\frac{1}{2}\cos^{2}\xi\mu^{i}\omega^{i}\,.
\end{align}

Since the solution corresponds to an AdS$_{3}$ vacuum, we can use the supersymmetry conditions defined in appendix D. of \cite{Conti:2023naw}. It is straightforward to check from the algebraic conditions given in that reference that the solution presented here is not supersymmetric.

\section{Holographic (Iso)spin from Isospin}\label{sec:Section4}

In this section we consider a fluctuation of the 10D dilaton, and show how the quantum numbers of the 2-sphere are mixed with the ones of the 3-sphere used to lift the lower dimensional gauged supergravity to 10D, mimicking the spin from isospin mechanism described in \cite{Jackiw:1976xx,Hasenfratz:1976gr}. 

\subsection{Scalar Fluctuation}

Since we are working in a pure 5-form sector of type II supergravity, in Einstein frame it is consistent to consider only dilaton fluctuations. The equation of motion for this fluctuation is just the 10D Laplacian
    \begin{equation}
        \nabla^{2}\Phi = \frac{1}{\ellAdS^{2}}\nabla^{2}_{\AdS{3}} \Phi +  g^{2} \nabla^{2}_{M_{7}}\Phi = 0,
    \end{equation}
where $\nabla^{2}_{\AdS{3}}$ is the Laplacian on AdS$_{3}$ of unit radius. We use a separation of variables ansatz
    \begin{equation}
        \Phi = \Phi_{\AdS{3}}\,\, \Phi_{\text{M}_{7}},
    \end{equation}
where $\Phi_{\AdS{3}}/\Phi_{\text{M}_{7}}$ has only support along AdS$_{3}$/M$_{7}$. Then, the 10D Laplacian amounts to solving two eigenvalue problems\footnote{Recall  from \eqref{eq:Constants5D}, $\ellAdS^{2} g^{2}=\frac{2}{7+\sqrt{5}}$.}
    \begin{equation}\label{eq:Laplacian7D}
        \nabla^{2}_{\AdS{3}}\Phi_{\AdS{3}} = \frac{2}{7+\sqrt{5}} M^{2}\Phi_{\AdS{3}}, \quad
        \nabla^{2}_{M_{7}}\Phi_{\text{M}_{7}} =-M^{2}\Phi_{\text{M}_{7}}.
    \end{equation}
The eigenvalues of the Laplacian of the internal manifold plays the role of a mass in the AdS$_{3}$ equation $m^{2} = \frac{2}{7+\sqrt{5}} M^{2}$. Then, in order for the solution to be stable, it should satisfy the Breitenlohner-Freedman (BF) bound, which for a scalar in AdS$_{d+1}$ reads $m^{2} \geq -d^{2}/4$. In this case, $M^{2}$ must satisfy
    \begin{equation}\label{eq:BFbound}
         M^{2} \geq -\frac{7+\sqrt{5}}{2}.
    \end{equation}

\subsection{Scalar Eigenfunctions of the \texorpdfstring{M$_{7}$}{M7} Laplacian}

The Laplacian on the 7D internal manifold can be written in terms of the Casimir elements of the isometry \eqref{eq:IsomM7} as
    \begin{equation}\label{eq:lapM7}
    \begin{aligned}
        \nabla^{2}_{\text{M}_7}\Phi &= \Lambda^{-2}(K^{i}K^{i}\Phi  -W^{2}\Phi) 
        + \frac{4}{\sin^{2}(\xi)}\partial^{2}_{\tau}\Phi
        + \frac{4}{\cos^{2}(\xi)} L^{i}L^{i}\Phi,\\
        &\phantom{=}
         + \frac{1}{\cos^{3}(\xi)\sin(\xi)}\partial_{\xi}\left( \cos^{3}(\xi)\sin(\xi) \partial_{\xi}\Phi  \right).
    \end{aligned}
    \end{equation}
At this level, solving the eigenvalues of the Laplacian is equivalent to the quantum mechanics problem of finding eigenstates of the Hamiltonian given by the Laplacian above. Such eigenstates are irreducible representations of the global symmetries of the Hamiltonian, which in this case are given by  \eqref{eq:IsomM7}. 

It is however important to recall that these SU(2) symmetries come from 3-spheres. For a round 3-sphere, with SO(4)$\simeq$SU(2)$_{L}\times$SU(2)$_{R}$ symmetry, the Laplacian eigenfunctions are labelled by the eigenvalues of the two Casimir elements of each SU(2), $\ell_{L}$ and $\ell_{R}$, and their respective magnetic moments, $m_{L}$ and $m_{R}$.  For scalar eigenfunctions we have $\ell_{L}=\ell_{R}=\ell$, which is equivalent to $L^{i}L^{i}=R^{i}R^{i}$. In what follows we denote by $(\ell,m_{L},m_{R})$ the labels on the 3-sphere $(\theta_{2},\varphi_{2},\psi)$ and $(\hat{\ell},\hat{m}_{L},\hat{m}_{R})$ the labels on the squashed 3-sphere $(\theta,\varphi,\tau)$.

In the case at hand, there is a mixing of symmetries between two 3-spheres. Since the SU(2)$_{L}$ symmetry of the 3-sphere ($\theta_{2},\varphi_{2},\psi$) is preserved, the eigenfunctions are labelled by $\ell$ and $m_{L}$. On the other hand, the U(1) symmetry generated by $\partial_{\tau}$ corresponds to the Cartan $\hat{R}^{3}$ element of the SU(2)$_{R}$ of the 3-sphere $(\theta,\varphi,\tau)$, this adds the label $\hat{m}_{R}$. The subtle point is the SU(2) generator of the SU(2)$_{D}$ symmetry. On one hand, its eigenstates must satisfy
    \begin{equation}\label{eq:eigenvaluesK}
        4 K^{i}K^{i} = - k(k+2), \quad K^{3} = i\, m_{k}.
    \end{equation}
However, since $K^{i} = \tilde{L}^{i}+R^{i}$, then $K^{i}K^{i} = R^{i}R^{i} + ...$, and since $R^{i}R^{i} = L^{i}L^{i}$, clearly $k$ is related to $\ell$. Similar argument applies for $m_{k}$ and $m_{R}$. We find that it is simpler to deal with $\tilde{L}^{i}$ and $R^{i}$ as if they where independent generators, and then impose that their eigenstates must satisfy \eqref{eq:eigenvaluesK}. This is similar to angular momentum coupling in quantum mechanics.

The solutions below are given in terms of eigenfunctions of the Laplacian on the two 3-spheres. These corresponds to the 3-sphere scalar spherical harmonics, which are given by
\begin{align}
Y^{\ell}_{m_{L},m_{R}}(\theta_{2},\varphi_{2},\psi) & =C_{\ell;m_{L},m_{R}}e^{\ri(m_{L}\psi+m_{R}\varphi_{2})}\cos^{m_{L}+m_{R}}\left( \frac{\theta_{2}}{2} \right) \sin^{m_{L}-m_{R}}\left(\frac{\theta_{2}}{2}\right)P_{\frac{\ell}{2}-m_{L}}^{(m_{L}-m_{R},m_{L}+m_{R})}(\cos\theta_{2})\,,\nonumber \\
Y^{\hat{\ell}}_{\hat{m_{L}},\hat{m_{R}}}(\theta,\varphi,\tau) & =C_{\ell;\hat{m_{L}},\hat{m_{R}}}e^{\ri(\hat{m_{L}}\tau+\hat{m_{R}}\varphi)}\cos^{\hat{m_{L}}+\hat{m_{R}}}\left( \frac{\theta}{2}\right)\sin^{\hat{m_{L}}-\hat{m_{R}}}\left(\frac{\theta}{2}\right)P_{\frac{\hat{\ell}}{2}-\hat{m_{L}}}^{(\hat{m_{L}}-\hat{m_{R}},\hat{m_{L}}+\hat{m_{R}})}(\cos\theta)\,,\nonumber \\
C_{\ell;m_{1},m_{2}} & =\sqrt{\frac{\ell+1}{2\pi^{2}}}\sqrt{\frac{(\frac{\ell}{2}+m_{1})!(\frac{\ell}{2}-m_{1})!}{(\frac{\ell}{2}+m_{2})!(\frac{\ell}{2}-m_{2})!}}\,.\label{familysolT}
\end{align}
where $P_{n}^{(\alpha,\beta)}(x)$ are the Jacobi polynomials satisfying
\begin{align}
(1-x^{2})\frac{\dd^{2}P_{n}^{(\alpha,\beta)}(x)}{\dd x^{2}}+(\beta-\alpha-(\alpha+\beta+2)x)\frac{\dd P_{n}^{(\alpha,\beta)}(x)}{\dd x}+n(n+\alpha+\beta+1)P_{n}^{(\alpha,\beta)}(x) & =0\,.
\end{align}
Each family of functions defined in (\ref{familysolT}) satisfies
\begin{align}
L^{i}L^{i}Y^{\ell}_{m_{L},m_{R}} & =-\frac{\ell(\ell+2)}{4}Y^{\ell}_{m_{L},m_{R}}\,,\\
\hat{L}^{i}\hat{L}^{i}\hat{Y}^{\ell}_{\hat{m}_{L},\hat{m}_{R}} & =-\frac{\hat{\ell}(\hat{\ell}+2)}{4}Y^{\hat{\ell}}_{\hat{m}_{L},\hat{m}_{R}}\,.
\end{align}
Regularity of the solutions require 
\begin{align}
\ell & =0,1,2,3,\dots\\
m_{L},m_{R} & \in\left\{ -\frac{\ell}{2}+n:n=0,1,2,\dots,\ell\right\} 
\end{align}

With this, we now proceed to the study of eigenfunctions of the Laplacian on M$_{7}$.

\subsubsection{Zero modes on the spheres}

The simplest case corresponds to $\Phi_{\text{M}_{7}} = \Phi_{\text{M}_{7}}(\xi)$. In this case, the equation for the dilaton reduces to
    \begin{equation}
        \frac{1}{\cos^{3}(\xi)\sin(\xi)}\partial_{\xi}\left( \cos^{3}(\xi)\sin(\xi) \partial_{\xi}\Phi  \right) = - M^{2} \Phi_{\text{M}_{7}},
    \end{equation}
with solution 
    \begin{equation}
    \begin{aligned}
        \Phi_{\text{M}_{7}} &= \frac{c_{1}}{z}\, \hypergeomF{-\frac{1}{2}\sqrt{4+M^{2}}}{\frac{1}{2}\sqrt{4+M^{2}}}{0}{\cos^{2}(\xi)} \\
        &\phantom{=}
        + c_{2}\,  \hypergeomF{1-\frac{1}{2}\sqrt{4+M^{2}}}{1+\frac{1}{2}\sqrt{4+M^{2}}}{2}{\cos^{2}(\xi)},
    \end{aligned}
    \end{equation}
were ${}_{2}F_{1}(a,b;c;z)$ is the ordinary hypergeometric function. Regularity of the solution at $\xi = 0$ and $\xi=\pi/2$ requires
    \begin{equation}
        c_{1}=0, \quad 
        M^{2} = 4(p^{2}-1), \quad \text{with } p=1,2,3,...
    \end{equation}
The mode $p=0$ is excluded since $\hypergeomF{1}{1}{2}{z} = -\log(1-z)/z$, which is singular at $z=1$ (here $z=\cos^{2}(\xi)$ so that $z=1$ is $\xi=\pi/2$).

Having determined $M^{2}$, we can explicitly check that the BF bound is satisfied since $M^{2}>0$ for any $p$.

\subsubsection{Spherical Harmonics on the squashed 3-sphere}

The next simplest subsector is to allow the dilaton to depend on $(\xi,\theta,\varphi,\tau)$. Since we are not exciting modes on the 3-sphere $(\theta_{2},\varphi_{2},\psi)$, effectively $K^{i} = \tilde{L}^{i}$ and $W = -\partial_{\tau}$. The equation now reads
    \begin{equation}
        \Lambda^{-2}\tilde{L}^{i}\tilde{L}^{i}\Phi+ \left( \frac{4}{\sin^{2}(\xi)} -\Lambda^{-2} \right) \partial^{2}_{\tau}\Phi + \frac{1}{\cos^{3}(\xi)\sin(\xi)}\partial_{\xi}\left( \cos^{3}(\xi)\sin(\xi) \partial_{\xi}\Phi  \right) =  -M^{2} \Phi_{\text{M}_{7}}
    \end{equation}
The first two terms correspond to the Laplacian on the squashed 3-sphere. We can then use a decomposition in spherical harmonics
    \begin{equation}
        \Phi = h(\xi) Y^{\hat{\ell}}_{\hat{m}_{L},\hat{m}_{R}}(\theta,\varphi,\tau) ,
    \end{equation}
which leads to the following equation for $h(\xi)$
    \begin{equation}
        -\frac{\Lambda^{-2}}{4}\hat{\ell}(\hat{\ell}+2)h-  \left( \frac{4}{\sin^{2}(\xi)} -\Lambda^{-2} \right) \hat{m}^{2}_{R}h + \frac{1}{\cos^{3}(\xi)\sin(\xi)}\partial_{\xi}\left( \cos^{3}(\xi)\sin(\xi) \partial_{\xi}h  \right) =  -M^{2} h
    \end{equation}
with solution
    \begin{equation}
    \begin{aligned}
        h(\xi) &= (-1)^{\hat{m}_{R}}(1-z)^{-\hat{m}_{R}} \hypergeomF{1-\hat{m}_{R}-p}{1-\hat{m}_{R}+p}{2}{\cos^{2}(\xi)},\\
        M^{2} &= 4(p^{2}-1) + \frac{1}{\Lambda^{2}}\left( \frac{1}{4}\hat{\ell}(\hat{\ell}+2)-\hat{m}_{R}^{2} \right),\quad
        p \geq |\hat{m}_{R}|+1,
    \end{aligned}
    \end{equation}    
where the condition $p \geq |\hat{m}_{R}|+1$ is imposed for regularity at the boundary. Again, all these fluctuations are above the BF bound.

\subsubsection{Solutions with angular momentum coupling}

We now study solutions with non-trivial dependencies on the two 3-spheres. A general solution of the eigenvalue problem \eqref{eq:Laplacian7D} is a linear combination of products of the functions in \eqref{familysolT}, with coefficients depending on $\xi$. For simplicity, let us consider a set of solutions of \eqref{eq:Laplacian7D} that depend on all the coordinates of the internal space but are annihilated by the operators
\begin{equation}\label{eq:Kto0}
K^{i}K^{i}\Phi =0,\qquad W\Phi =0\,.
\end{equation}
In the quantum mechanical problem, this is equivalent to finding a set of pairs of states of quantum numbers $\ket{\ell,m}\otimes\ket{\hat{\ell},\hat{m}}$ such that the total angular momentum of the system (the $K^{i}$ quantum numbers) are $\ket{k=0,m_{k}=0}$.

It is possible to find a set of functions satisfying \eqref{eq:Kto0} labelled by $\ell,m_{R},\hat{m}_{L}$. Explicitly, they are given by
\begin{align}
\Phi_{0}^{\ell,m_{R},\hat{m}_{L}} & =\sum_{n=-\ell/2}^{\ell/2}e^{-\frac{\ri\pi m_{R}}{2}}Y^{\ell}_{n,m_{R}}(\theta_{2},\varphi_{2},\psi)Y^{\ell}_{\hat{m}_{L},n}(\theta,\varphi,\tau)\,.
\end{align}
These functions also satisfy
\begin{align}
L^{i}L^{i}\Phi_{0}^{\ell,m_{R},\hat{m}_{L}} & =-\frac{\ell(\ell+2)}{4}\Phi_{0}^{\ell,m_{R},\hat{m}_{L}}\,,\\
\partial_{\tau}^{2}\Phi_{0}^{\ell,m_{R},\hat{m}_{L}} & =-\hat{m}_{L}^{2}\Phi_{0}^{\ell,m_{R},\hat{m}_{L}}\,.
\end{align}
Let us consider $\Phi_{M_7}=h(\xi)\Phi_{0}^{\ell,m_{R},\hat{m}_{L}}$. Then, the eigenvalue problem for the M$_{7}$ Laplacian reads
\begin{align}
\frac{1}{\cos^{3}\xi\sin\xi}\partial_{\xi}(\cos^{3}\xi\sin\xi\partial_{\xi}h)-\frac{4\hat{m}_{L}^{2}}{\sin^{2}\xi}h-\frac{\ell(\ell+2)}{\cos^{2}\xi}h & =-M^{2}h\,.
\end{align}
The solution of this equation is given by
\begin{align}
h(\xi) & =\frac{1}{\sin^{2\hat{m}_{L}}(\xi)}\left[\frac{C_{1}}{\cos^{2+\ell}(\xi)}\,_{2}F_{1}\left(a-1-\ell,b-1-\ell;-\ell;\cos^{2}\xi\right)+\right.\nonumber \\
 & \hspace{4em}\hspace{2em}\left.\phantom{\frac{1}{2}}+C_{2}\cos^{\ell}(\xi)\,_{2}F_{1}\left(a,b;\ell+2;\cos^{2}\xi\right)\right]\\
a & \equiv1+\frac{\ell}{2}-\sqrt{1+\frac{M^{2}}{2}}-2\hat{m}_{L}\,,\\
b & \equiv1+\frac{\ell}{2}+\sqrt{1+\frac{M^{2}}{2}}-2\hat{m}_{L}\,.
\end{align}
where $C_1,C_2$ are integration constants. Let us analyse the regularity at $\xi=0,\frac{\pi}{2}$. Let $\Gamma(x)$
be the Euler Gamma function. For any $_{2}F_{1}(a,b,c;\cos^{2}\xi)$, the expansion closed to $\xi=0$ goes as
\begin{align}
_{2}F_{1}\left(a,b;c;\cos^{2}\xi\right) & \simeq\Gamma(c)\left(\frac{\pi}{90\sin(\pi(a+b-c))\Gamma(1+a+b-c)\Gamma(-a+c)\Gamma(-b+c)}+\mathcal{O} (\xi^{2})\right)\,,\nonumber \\
 & \text{for }-a-b+c>0\,. \label{solhypertutto}
\end{align}
where $\mathcal{O}(\xi^{2})$ are subleading terms that do not contain $\Gamma(c)$. The hypergeometric function controlled by $C_{1}$ in \eqref{solhypertutto} has coefficient $c=-\ell$, thus, this branch is ill defined at $\xi=0$. We set $C_{1}=0$. The branch controlled by $C_{2}$ has the following expansion
\begin{align*}
\xi^{-2\hat{m}_{1}}&\left(\frac{\pi\Gamma(2+\ell)}{\Gamma(-1+a+b-\ell)\Gamma(2-a+\ell)\Gamma(2-b+\ell)}+\mathcal{O}(\xi^{2})\right)- \\
&\hspace{3cm} -\xi^{2\hat{m}_{1}}\left(\frac{\pi\Gamma(2+\ell)}{\Gamma(a)\Gamma(b)\Gamma(3-a-b+\ell)}+\mathcal{O}(\xi^{2})\right)
\end{align*}
Since $\hat{m}_{1}\in\{-\ell/2,\dots,\ell/2\}$, let us split the analysis in three cases:
\begin{itemize}
    \item For $\hat{m}_{1}>0$, then regularity can be
    obtained by imposing
        \begin{align}
        2-a+\ell & =-p\,,\qquad p=0,1,2,\dots
        \end{align}
    which is solved by the quantization condition
        \begin{align}
        M^{2} & =\ell^{2}+4\ell(1+p+\hat{m}_{1})+4[\hat{m}_{1}^{2}+2\hat{m}_{1}(1+p)+(2+p)p]\,.
        \end{align}
    \item For $\hat{m}_{1}<0$, regularity at $\xi =0 $ is achieved by imposing $a=-p$, which leads to the quantization condition
        \begin{align*}
        M^{2} & =\ell^{2}+4\ell(1-\hat{m}_{1}+p)+4(\hat{m}_{1}^{2}-2\hat{m}_{1}(1+p)+p(2+p))\,,\qquad\hat{m}_{1}<0\,.
        \end{align*}

    \item For $\hat{m}_{1}=0$ (which is only possible for even $\ell$) a logarithmic branch appears, which can be eliminated by imposing
    \begin{align*}
    M^{2} & =\ell^{2}+4\ell(1+p)+4p(p+2)\,,
    \end{align*}
    leading to constant behaviour near $\xi=0$. 
\end{itemize}
Regularity at $\xi=\pi/2$ does not requires any extra condition. In all the cases discussed above we have $M^{2}>0$, hence, we are always above the BF bound.

\section{Conclusion}\label{sec:Conclusions}

Motivated by the global-gauge symmetry mixing produced by the presence of  non-Abelian hedgehog monopoles in 4D SU(2) gauge theories discussed in  \cite{Jackiw:1976xx,Hasenfratz:1976gr}, in this paper we reviewed a particular SU(2) non-Abelian monopole configuration known as Meron gauge fields. We showed that for this configuration, the argument of \cite{Jackiw:1976xx,Hasenfratz:1976gr} applies, and also reviewed generalities of uplifts of this type of gauge fields using 3-spheres. At this level, the global-gauge symmetry mixing becomes a diagonal combination of the isometries of the 2-sphere where the non-Abelian monopole is valued, and the 3-sphere used to lift the configuration. We also find an additional Killing vector that generates a U(1) isometry.

We studied two examples of Type II configurations that have the features described above. The first case, corresponds to a deformation of the 3D theory known as I-branes. The background corresponds to formulating the theory on a 2-sphere with a Meron monopole, and it was obtained in \cite{Canfora:2021nca} at the level of gauged supergravity. The background corresponds to a 2-dimensional black hole studied in \cite{Mann:1991ny,Lemos:1994fn,Witten:1991yr} and an 8-dimensional internal manifold, formed by a 2-sphere fibered over two 3-spheres by means of the Meron gauge field. For vanishing black hole mass the configuration is supersymmetric. We find that the Killing spinor is charged under the diagonal symmetry. In the supersymmetric limit, the metric is regular everywhere but the 10D dilaton is divergent in an asymptotic region. The origin of this singularity can be traced down to the fact that the configuration is constructed out of NS5-branes.

The second configuration is new and it corresponds to a solution of 5D SU(2)$\times$U(1) gauged supergravity. The metric is a direct product of the form AdS$_{3}\times\Sp{2}$, while the SU(2) gauge field is a Meron and there is an Abelian monopole for the U(1) gauge field. When uplifted to Type IIB using \cite{Lu:1999bw}, the background corresponds to a deformation of AdS$_{5}\times\Sp{5}$, where the D3-branes are wrapping a 2-sphere. In this case, the lower dimensional 2-sphere and one of the $\Sp{5}$ U(1) directions can be though as (locally) a squashed 3-sphere whose fiber shrinks. There is a second 3-sphere which the squashed one is fibered over. The diagonal symmetry in this case mixes the SU(2)$_{L}$ of the squashed 3-sphere with the SU(2)$_{R}$ of second 3-sphere. This AdS$_{3}$ configuration is not supersymmetric for any range of the parameters and it is regular everywhere. 

Finally, we studied dilaton fluctuations on the uplift of the  AdS$_{3}\times\Sp{2}$ background. The equation of motion for the fluctuation reduced to an eigenvalue equation for the Laplacian on AdS$_{3}$ and M$_{7}$, where M$_{7}$ corresponds to the 2-sphere fiber over the 5-sphere used to uplift the geometry to 10-dimensions. We find that for all fluctuations, the effective AdS$_{3}$ mass is above the Breitenlohner-Freedman bound, so that the non-supersymmetric background is stable with respect to dilaton fluctuations. On M$_{7}$, the eigenfunctions are labeled by quantum numbers of the global symmetries of the manifold. We find fluctuations which show angular momentum mixing between the SU(2) spins of two 3-spheres inside M$_{7}$, mimicking the spin from isospin mechanism of \cite{Jackiw:1976xx,Hasenfratz:1976gr}. 

Going back to the AdS$_{3}\times\Sp{2}$ solution, from the dual 2D CFT perspective, the diagonal SU(2) symmetry between the SO(3) global and SU(2) gauge symmetries corresponds to an symmetry of the internal space. This is to be compared with \cite{Jackiw:1976xx,Hasenfratz:1976gr} where the symmetry being mixed is a global (Lorentz) symmetry. Due to this, the diagonal symmetry does not produce a shift of the spin, in the sense of representations of the Lorentz group, but rather a shift of the SO(3) isospin, so that we instead have an ``isospin from isospin" mechanism. Furthermore,  in \cite{Jackiw:1976xx,Hasenfratz:1976gr} the global symmetry is mixed with a gauge one, whereas here the internal symmetry is mixed with a second internal symmetry. In any case, these results serve as a first approach to obtaining a background which truly realises the results of \cite{Jackiw:1976xx,Hasenfratz:1976gr}.

From here, there are several interesting directions to explore

\begin{itemize}
    \item We note that the 5D SU(2) $\times$ U(1) gauged supergravity can also be uplifted to the infinite family of Gaiotto-Maldacena solutions \cite{Gaiotto:2009gz} using \cite{Chatzis:2024kdu,Chatzis:2024top}. Due to our construction, it is guaranteed that there will be a diagonal symmetry between the 2-sphere from the gauged supergravity, and the 2-sphere of the internal manifold of the Gaiotto-Maldacena solutions. This shows the universality of this phenomena. It would be interesting to study this uplift as a non-supersymmetric reduction of 4D $\mathcal{N}=2$ SCFTs.
    \item One possibility to obtain a solution realising a diagonal symmetry between a Lorentz symmetry and a internal symmetry, is to interpret the AdS$_{3}$ solution presented here as a IR fixed point of a more general background of the form
        \begin{equation}
            \ds_{5D} = e^{2f(r)}\dd x^{2}_{1,1} + e^{2g(r)}\ds(\text{S}^{2}) + \dd r^{2}.
        \end{equation}
    If this solution exists, from the point of view of a 4D observer, there would really be a spin from isospin effect. A more general profile for the SU(2) and U(1) gauge fields and a non-zero dilaton might be necessary.
    \item It would be interesting to further study solutions containig a Meron gauge field, for example the Meron gauge field on a D5-brane wraped on a 2-sphere of \cite{Nunez:2023xgl}.
    \item We showed in Section \ref{sec:Section2} that the lifts studied here have an additional U(1) isometry. At the level of gauge supergravity, this isometry is related to the generator $\mu^{i}T^{i}$. It would be interesting further analyse this connection.
    \item The key ingredient in realizing the diagonal symmetry is that the manifolds $\text{S}^2$ and $\text{S}^3$ share the same isometry algebra. A possible generalization is to consider manifolds that share a common subalgebra of their isometry groups, allowing one to construct 1-forms whose Lie derivatives along this subalgebra close among themselves. In principle, this would allow the construction of manifolds with non-trivial fibrations that can serve as internal spaces for solutions of 10- or 11-dimensional supergravity.
    \item It would be interesting to pursue a further stability analysis in the non-supersymmetric AdS$_{3}$ solution. Here, we took advantage of the fact that in the purely metric and $F_{5}$ sector of Type IIB, it is consistent to study dilaton perturbations. The mass of these fluctuations, dual to spin-0 glueballs, are always above the BF bound, so a instability is not triggered by this fluctuations. A similar analysis where a set of fluctuations decouple from rest can be performed to the spin-2 fluctuation studied in \cite{Bachas:2011xa}. A more general study stability requires a systematic perturbative analysis. 
\end{itemize}

We hope to report on these and related topic in the future.

\section*{Acknowledgements}

We thank Evangelos Afxonidis, Adolfo Guarino, Yolanda Lozano, Niall T. Macpherson, Carlos Nunez, Mario Trigiante for valuable discussions. The work of M.O.   is partially funded by Beca ANID de Doctorado grant 21222264. The work of R. S. is supported by the Ram\'on y Cajal fellowship RYC2021-033794-I, and by grants from the Spanish government MCIU-22-PID2021-123021NB-I00 and principality of Asturias SV-PA-21-AYUD/2021/52177.


\bibliographystyle{JHEP}

\bibliography{ref}

\end{document}